# Adaptive Market Efficiency of Agricultural Commodity Futures Contracts

Semei Coronado Ramírez[1]
Pedro Luis Celso Arellano[2]
Omar Rojas[3]

**Abstract**

In this paper we investigate the adaptive market efficiency of the agricultural commodity futures market, using a sample of eight futures contracts. Using a battery of nonlinear tests, we uncover the nonlinear serial dependence in the returns series. We run the Hinich portmanteau bicorrelation test to uncover the moments in which the nonlinear serial dependence, and therefore adaptive market efficiency, occurs for our sample.

## Eficiencia del Mercado Adaptativo en los Contratos Futuros de Productos Agrícolas

**Resumen**

En este documento se investiga la eficiencia del mercado adaptativo del mercado de futuros de productos básicos agrícolas, utilizando una muestra de ocho contratos de futuros. Se utiliza una batería de pruebas no lineales para descubrir la dependencia no lineal en la serie de retornos. Aplicamos el estadístico Hinich portmanteau bicorrelación para descubrir los momentos de dependencia no lineal en las series, y por lo tanto se encuentra que cuatro productos del mercado tienen adaptable eficiencia

Palabras claves: mercados eficientes, no linealidad, hipótesis de mercados adaptativos, productos agrícolas, mercado de futuros

## Introduction

The weak-form efficient market hypothesis (EMH) is a financial theory that has attracted lots of attention from researchers for over four decades. A market is efficient when prices always fully reflect available information (Fama, 1970). However, despite such a large body of research on EMH, there is no consensus on whether markets are efficient or not. Thus, according to Campbell *et al.* (1997), the notion of relative efficiency may be a more useful concept than the all-or-nothing view taken by the conventional efficiency studies. They suggest relative efficiency because measuring efficiency provides more insights than testing it, *i.e.*, it may be more useful to know the differences in the degree of inefficiency across markets than knowing that a market is inefficient *per se*.

[1] Universidad de Guadalajara, Profesor del Departamento de Métodos Cuantitativos, México; e-mail: semeic@cucea.udg.mx

[2] Universidad de Guadalajara, Profesor del Departamento de Sistemas de Información, México; e-mail: pcelso@cucea.udg.mx

[3] Universidad Panamericana, Campus Guadalajara, Profesor-Investigador de la Escuela de Ciencias Económicas y Empresariales, México; e-mail: orojas@up.edu.mx

Traditionally, the weak-form EMH has been tested in empirical studies through the unpredictability of returns from past returns criterion (or conventional efficiency studies). Some of these tools are: the serial autocorrelation test using the Ljung and Box (1978) portmanteau Q statistic, the runs test (Shiller and Perron, 1985) and the variance ratio test (Al-Khazali *et al.*, 2007; Chow and Denning, 1993; Lo and MacKinlay, 1988), amongst others. However, these short-horizon return predictability studies have been criticized for their focus on linear correlations of price changes. According to Hong and Lee (2005), an alternative approach would be to remove all linear serial autocorrelation from the sample and determine whether returns still contain predictable nonlinearities. Furthermore, since a white noise process does not necessarily imply efficiency as returns series can be linearly uncorrelated and at the same time nonlinearly dependent (Granger, 2001; Granger & Andersen, 1978), there is a need for the utilization of nonlinear techniques to uncover hidden nonlinear serial dependency structures in time series, *cf.* (Hinich and Patterson, 1985; Hsieh, 1991; Panagiotidis and Pelloni, 2007; Patterson and Ashley, 2000). Amongst the vast number of nonlinear tests, there is the McLeod and Li (1983) Test, Tsay (1986) Test, ARCH-LM Test (Engle, 1982) and the BDS Test (Broock *et al.*, 1996). All these tests can be used to detect the presence of nonlinear dependence in a time series and therefore market inefficiency.

An alternative to the EMH from a behavioral perspective, the adaptive market hypothesis (AMH), proposed by Lo (2004), states that markets are adaptable and switch between efficiency and inefficiency at different epochs. In this theory (Lo, 2005), the degree of market efficiency is related to environmental factors characterizing market ecology such as the number of competitors, the magnitude of profit opportunities available and the adaptability of market participants. Some practical implications of the

AMH are: there are changes over time in the risk-reward relationships due to the preferences of the market population; current preferences are influenced by the movement of past prices due to the forces of natural selection, in contrast to the weak-form of EMH where history of prices is not taken into account; arbitrage opportunities, being constantly created and disappearing, exist at different points in time.

Under the AMH point of view, it is desirable to detect nonlinear phenomena in certain periods of time and not only in the full series. For this purpose, it can be of use the framework originally proposed by Hinich and Patterson (1995), then published as (Hinich and Patterson, 2005), in which the full sample period is divided into equal-length non-overlapped moving time windows, and the Hinich (1996) Portmanteau Bicorrelation Test Statistic is computed for detecting nonlinear serial dependence in each window and therefore inefficiency. The aforementioned test has been successfully applied to analyze the nonlinear behavior of different financial and economic time series, *cf.* (Bonilla *et al.*, 2008; Coronado-Ramírez and Arreola, 2011; Hinich and Serletis, 2007; Lim *et al.*, 2008; Romero-Meza *et al.*, 2007).

Recently, a number of researchers have shifted from the traditional focus of absolute and static EMH to tracking the changing degree of efficiency over time, giving way to the AMH point of view: Hiremath and Kumari (2014) studied India´s stock market; Alvarez-Ramirez *et al.* (2012), Ito and Sugiyama (2009) and Kim *et al.* (2011) found time varying efficiency in the U.S; Charles *et al.* (2012) and Neely *et al.* (2009) gave evidence of AMH in the foreign exchange market; Noda (2012) focused on Japan, Lim *et al.* (2008) on Asian Markets and Urquhart and Hudson (2013) on some major stock markets; as for commodity markets, Coronado-Ramírez *et al.* (2014) explored the international coffee market, applying nonlinear statistical tests to detect periods of inefficiency for the case of Colombian Arabica beans.

In this paper we focus our attention on the futures market. In particular, we study the nonlinear serial behavior of the returns of eight agricultural commodity futures contracts traded at the Chicago Mercantile Exchange (CME), namely: CBOT corn, KC HRW Wheat, CBOT Soybean Oil, ICCO Cocoa, ICO Coffee Arabica Mild Average, ISA Raw Sugar, Feeder Cattle and Eggs Large White. We have chosen these commodities since they are some of the most traded and representative of the agricultural commodity futures markets traded at the CME, the leader in global marketplace. All the prices were collected from *Thomson Reuters Datastream Professional* for the sample period 7/7/1994 to 11/15/2010, for a total of 4267 observations for each series. The selected time frame allows for a comprehensive study since the sample takes into account periods of high and low volatility, including the sub-prime global financial crisis.

Understanding the underlying core behavior of a return series of commodity futures is vital for better decision making for producers, investors, traders and policy makers (Karali & Power, 2009). Agricultural economic time series have historically been modeled to predict prices and/or volatility of these products (commodities) with autoregressive (AR) models, moving average (MA), autoregressive integrated moving average (ARIMA), transfer functions and dynamic analysis (Aradhyula & Holt, 1988). Also, in recent times, generalized autoregressive heteroskedastic (GARCH) models have been used (Adrangi & Chatrath, 2003; Benavides, 2004; Tansuchat *et al.*, 2009), as well as smooth transition vector error correction models (STVECM) (Milas & Otero, 2002). There are other studies that show the chaotic behavior in the prices of commodities, these include Lyapunov exponents test, the Brock Dechert and Scheinkman statistic, (BDS), the correlation exponent, neural networks, amongst others (Ahti, 2009; Blank, 1991; Tejeda & Goodwin, 2009; Velásquez & Aldana, 2007; Yang & Brorsen, 1993).

The present study aims to complement and extend existing work on AMH, uncovering nonlinear serial dependence not only on the full sample, using the nonlinear techniques mentioned, but also exhibiting windows of time in which such phenomena occur, with the aid of the Hinich test. To the best of our knowledge, no AMH point of view has been applied to commodity futures contracts using these tools.

The plan of this paper is as follows: Section 2 describes the methodology used. Section 3 describes the data. Section 4 reports the empirical results. Concluding remarks are presented at the end of the paper.

**Methodology**

In this section we present the battery of nonlinear tests applied to uncover nonlinear serial dependence and empirically test the EMH and AMH of our sample. Even if we focus mainly on showing how nonlinear phenomena is present in certain periods of time, we apply such a battery of tests in order to present a more robust study and avoid the risk of overemphasizing the generality of the findings. We use the tests already mentioned in the introduction, namely: McLeod-Li test, Tsay test, ARCH-LM test, BDS test and Hinich test. Before running them on the pre-whitened returns series, we perform two unit root tests to check for stationarity of the returns series. Besides running the traditional Augmented Dickey Fuller Test, which assumes normality of errors (Dickey and Fuller, 1979), we test for stationarity using the Residual Augmented Least Squares (RALS) methodology, which does not require knowledge of a specific density function or a functional form (Im, Lee and Tieslau, 2014).

The nonlinear tests employed are of two kinds, all of which are such that rejection of the null hypothesis implies the presence of nonlinear dependence in the series and, therefore, market inefficiency. The fist kind of tests we run are of static and absolute nature, meaning that they apply for the entire returns series; the second kind, which is the main focus and contribution of our paper, is a moving time non-overlapped windows approach, which tests for nonlinearity in different periods of time and thus gives evidence of the AMH. Prior to performing these tests, data pre-whitening is necessary to remove any linear structure from the data, so that any remaining serial dependence is due to nonlinear phenomena. The linear dependence is removed form the sample by fitting an autoregressive model of order $p$, $AR(p)$, where the optimal lag is chosen to minimize the Schwarz's Bayesian Information Criterion. These white noise residuals are thus analyzed with the following tests: the McLeod-Li test of nonlinearity which is used to find out if the squared autocorrelation function of returns is non-zero (McLeod and Li, 1983); the Tsay test, used to detect the quadratic serial dependence in the sample (Tsay, 1986); the ARCH-LM test used to detect ARCH distributive (Engle, 1982); the BDS portmanteau test for time-based dependence in the series (Broock et al., 1996); finally, the Hinich portmanteau bicorrelation test (Hinich, 1996), which is a third order extension of the standard correlation tests for white noise and detects nonlinear serial dependence in non-overlapped time windows. Since this is the method employ to give evidence of the AMH on the series under study, we explain it in more detail in the following subsection.

*The Hinich portmanteau bicorrelation test*

We now proceed to explain the windowed test procedure used in this study, the Hinich portmanteau bicorrelation test (*H-test*). Such a test is used to detect epochs of transient dependence in a discrete-time pure white noise process and involves a procedure of dividing the full sample period into equal length non-overlapping moving time windows or frames on each of which the portmanteau bicorrelation statistic (*H-statistic*) is computed, to detect nonlinear serial dependence. Let the sequence $\{x(t_k)\}$ denote the sampled data process at a fixed rate, where the time unit $t$ is discrete. The H-test employs non-overlapped time windows, thus if we denote by $n$ the window length, then the $k$-th window is $\{x(t_k), x(t_k + 1), \ldots, x(t_k + n - 1)\}$. The next window is $\{x(t_{k+1}), x(t_{k+1} + 1), \ldots, x(t_{k+1} + n - 1)\}$, where $t_{k+1} = t_k + n$. All observations are standardized $z(t_k) = (x(t_k) - \mu_x)/\sigma_x$, where $\mu_x$ and $\sigma_x$ are the expected value and the standard deviation of each process, respectively. The null hypothesis for each window is that $\{z(t_k)\}$ are realizations of a stationary pure white noise process that has zero bicorrelation, defined by $C_{zzz}(r,s) = E[z(t)z(t+r)z(t+s)]$ for $0 < r < s < L$, where $L$ is the number of lags in each window. The alternative hypothesis is that the process generated in the window is random with some non-zero bicorrelations, *i.e.*, there exists third-order nonlinear dependence in the data generation process.

The H-statistic, used to detect nonlinear dependence within a window, and its corresponding distribution, are

$$H := \sum_{s=2}^{L} \sum_{r=1}^{s-1} \frac{G^2(r,s)}{T-s} \sim \chi^2_{L(L-1)/2}$$

where $G(r,s) = (n-s)^{1/2} C_{zzz}(r,s)$. The number of lags $L$ is specified as $L := n^c$, with $0 < c < 0.5$, where $c$ is a parameter to be chosen. To maximize the power of the test and ensure the asymptotic properties, Hinich and Patterson (1995, 2005) suggest, based on results from Monte Carlo simulations, to set $c = 0.4$. A window will be statistically significant if the null hypothesis is rejected at the specified threshold level set to 0.05.

*The data*

For the present study we consider daily returns of eight agricultural commodity futures prices traded at the Chicago Mercantile Exchange (CME) of the following products: CBOT corn, KC HRW Wheat, CBOT Soybean Oil, ICCO Cocoa, ICO Coffee Arabica Mild Average, ISA Raw Sugar, Feeder Cattle and Eggs Large White. All the prices of these futures contracts were collected from *Thomson Reuters Datastream Professional* for the sample period 7/7/1994 to 11/15/2010, for a total of 4267 observations for each of the 8 futures contracts. These dates were chosen in such a way that all contracts had observations in these periods. Prices where transformed into series of continuously compounded percentage returns by taking the differences of the logarithm of the prices, i.e. $r_t = \ln(p_t) - \ln(p_{t-1})$, where $p_t$ is the price of the futures contract on day $t$. Figure 1 and figure 2 show the behavior of the prices and returns for each series.

**Figure 1**

**Prices (upper graphs) and the corresponding returns (lower graphs) for Corn, Wheat, Cocoa and Soybean Oil futures contracts, from left to right, respectively**

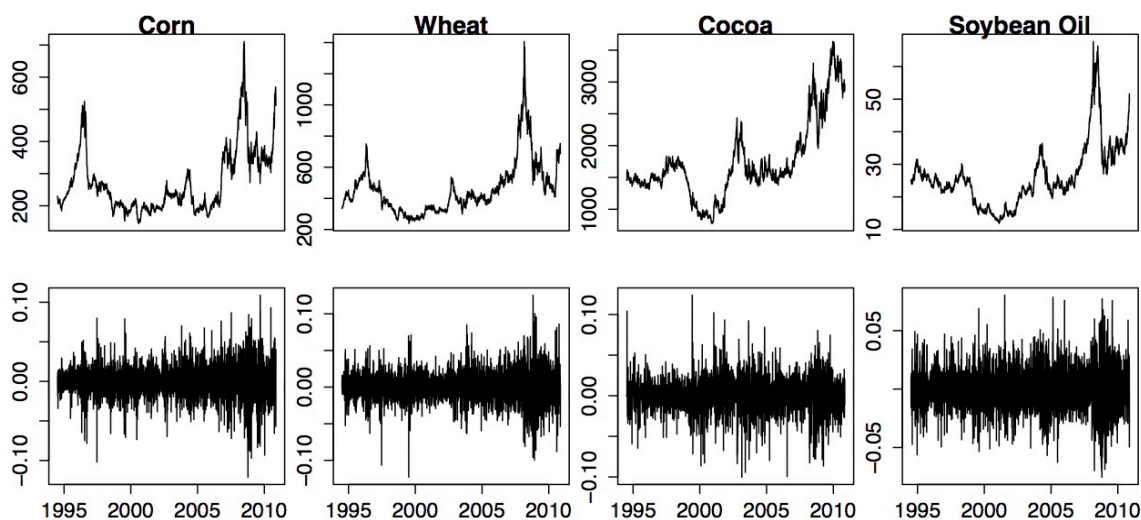

## Figure 2

**Prices (upper graphs) and the corresponding returns (lower graphs) for Sugar, Coffee, Feeder Cattle and Eggs futures contracts, from left to right, respectively**

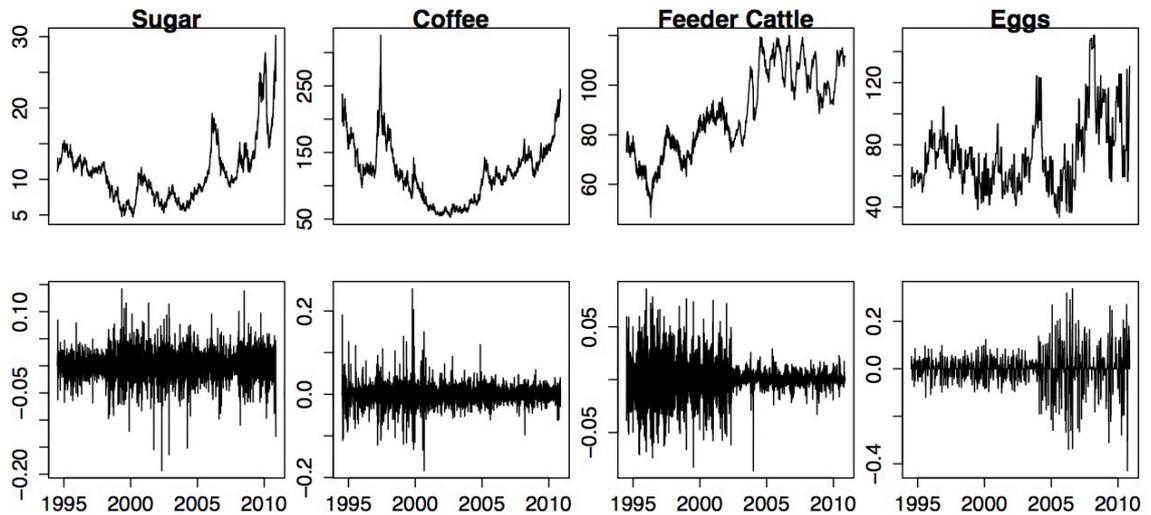

**Empirical results**

In this section we discuss the empirical results of stationarity and nonlinear tests carried out. Summary statistics for the returns are shown in table 1. The statistics are consistent, as expected, with some of the stylized facts of financial time series (Cont, 2001). In particular, the kurtosis indicates that return distributions are leptokurtic. Furthermore, the Jarque and Bera (1987) statistic (JB) confirms returns not normally distributed.

## Table 1
**Summary statistics for the returns of the futures contracts**

|  | Mean | Min | Max | StdDev | Skewness | Kurtosis | JB |
|---|---|---|---|---|---|---|---|
| **Corn** | 0.0002 | -0.121 | 0.109 | 0.018 | -0.190 | 6.616 | 2351 |
| **Wheat** | 0.0002 | -0.123 | 0.125 | 0.019 | 0.055 | 7.014 | 2867 |
| **Cocoa** | 0.0002 | -0.100 | 0.124 | 0.017 | -0.122 | 8.925 | 6253 |
| **Soybean Oil** | 0.0002 | -0.075 | 0.080 | 0.016 | 0.095 | 5.175 | 847.7 |
| **Sugar** | 0.0002 | -0.193 | 0.142 | 0.022 | -0.364 | 9.822 | 8367 |
| **Coffee** | 0.0000 | -0.184 | 0.253 | 0.021 | 0.476 | 21.030 | 57987 |
| **Feeder Cattle** | 0.0000 | -0.086 | 0.086 | 0.015 | 0.014 | 7.738 | 3991 |
| **Eggs** | 0.0002 | -0.429 | 0.336 | 0.037 | -0.834 | 30.770 | 137611 |

*Stationarity tests*

In table 2, we present results of the ADF and RALS tests. As it is well known, as more negative the value of the statistic, the stronger the rejection of the hypothesis that there is a unit root at the 5% significance level chosen. As we can observe, both tests imply rejection of the null hypothesis.

**Table 2**
**Results from the ADF and RALS stationarity tests**

|  | ADF | RALS |
|---|---|---|
| **Corn** | -14.869 | -16.569 |
| **Wheat** | -18.465 | -19.555 |
| **Cocoa** | -17.833 | -19.638 |
| **Soybean Oil** | -16.991 | -18.636 |
| **Sugar** | -18.992 | -19.895 |
| **Coffee** | -17.882 | -18.479 |
| **Feeder Cattle** | -14.870 | -15.623 |
| **Eggs** | -20.764 | -23.480 |

Significant at 5%. The critical value for the ADF test is -2.87

*Tests of nonlinear serial dependence*

In this subsection we present the results of the battery of nonlinear tests applied to the sample. A series of conventional linear tests, as those mentioned in the introduction, were run beforehand and are available from the authors upon request. These tests are incapable of capturing nonlinear patterns in the data. As is well known, the failure to reject linear dependence is insufficient to prove independence of the series when in presence of non-normality (Hsieh, 1991) and not necessarily imply independence (Granger and Andersen, 1978). Given that the presence of nonlinearity in the return series contradicts the EMH, we employ the nonlinear tests to investigate nonlinear dependence.

Most of the series were adjusted between an *AR(1)* and an *AR(4)*, prior to the application of the tests, in order to remove any linear dependence, where the optimal lag was chosen to minimize the Schwarz's Bayesian Information Criterion. Therefore, we reject the null hypothesis, given the significance level of 5% (see table 3), where the pure white noise is explained solely by the nonlinearity of the historic series. The results in table 3 as a whole provide evidence of nonlinearity in all the returns series, though "unanimous" verdict from all the tests is reached for all but the Cocoa time series, in which the Tsay test using lags up to 5 and the BDS test (m=2) cannot reject the null hypothesis at the 5% level of significance.

**Table 3**
**Nonlinearity test results**

|  | Corn | Wheat | Cocoa | Soybean Oil | Sugar | Coffee | Feeder Cattle | Eggs |
|---|---|---|---|---|---|---|---|---|
|  | AR(2) | AR(2) | AR(4) | AR(2) | AR(1) | AR(2) | AR(2) | AR(1) |
| McLeod-Li test |  |  |  |  |  |  |  |  |
| Lag 5 | 0.000 | 0.000 | 0.000 | 0.000 | 0.000 | 0.000 | 0.000 | 0.000 |
| Lag 15 | 0.000 | 0.000 | 0.000 | 0.000 | 0.000 | 0.000 | 0.000 | 0.000 |
| Lag 20 | 0.000 | 0.000 | 0.000 | 0.000 | 0.000 | 0.000 | 0.000 | 0.000 |
|  |  |  |  |  |  |  |  |  |
| Tsay test |  |  |  |  |  |  |  |  |
| Lag 5 | 0.030 | 0.002 | 0.870 | 0.001 | 0.000 | 0.000 | 0.000 | 0.000 |
| Lag 15 | 0.000 | 0.000 | 0.002 | 0.000 | 0.000 | 0.000 | 0.000 | 0.000 |
| Lag 20 | 0.000 | 0.000 | 0.001 | 0.000 | 0.000 | 0.000 | 0.000 | 0.000 |
|  |  |  |  |  |  |  |  |  |
| Engle LM test |  |  |  |  |  |  |  |  |
| Lag 5 | 0.000 | 0.000 | 0.001 | 0.000 | 0.000 | 0.000 | 0.000 | 0.000 |
| Lag 15 | 0.000 | 0.000 | 0.000 | 0.000 | 0.000 | 0.000 | 0.000 | 0.000 |
| Lag 20 | 0.000 | 0.000 | 0.000 | 0.000 | 0.000 | 0.000 | 0.000 | 0.000 |
|  |  |  |  |  |  |  |  |  |
| BDS test |  |  |  |  |  |  |  |  |
| $m = 2, \varepsilon = 0.5s$ | 0.000 | 0.000 | 0.159 | 0.000 | 0.000 | 0.000 | 0.000 | 0.000 |
| $m = 3, \varepsilon = s$ | 0.000 | 0.000 | 0.009 | 0.000 | 0.000 | 0.000 | 0.000 | 0.000 |
| $m = 4, \varepsilon = 1.5s$ | 0.000 | 0.000 | 0.001 | 0.000 | 0.000 | 0.000 | 0.000 | 0.007 |
|  |  |  |  |  |  |  |  |  |
| Hinich bispectrum test | 0.000 | 0.000 | 0.000 | 0.000 | 0.000 | 0.000 | 0.000 | 0.000 |

Significant at 5%. For the BDS test, m and ε denote the embedding dimension and distance, respectively and ε is equal to various multiples of the sample standard deviation

*Test of nonlinearity for the AMH*

In spite of the fact that the results of the preceding subsection gave strong evidence of nonlinearity in all the returns series, it is actually possible that the significant results are driven by nonlinear phenomena within a small number of periods. Such periods would be indicative of evidence of AMH on the commodity futures market, for the sample under study. For this purpose, we perform an H-test on 152 non-overlapped moving time windows of 28 daily observations, to detect possible nonlinear serial dependence in each window. Having removed the possibility of spurious autocorrelations by the pre-whitening of the series, Table 4 presents the results of the H-test, where the number of significant windows, indicating the presence of nonlinear serial dependence is identified, along with the epochs in which it occurs. These results are consistent with other works, which have applied the H-test to financial or economic time series to detect nonlinear dependence periods (Hinich and Serletis 2007; Lim *et al.*, 2008, Bonilla *et al.*, 2008; Coronado and Gatica, 2011).

An interesting insight from the Hinich framework is that in this way it is possible to assess the relative efficiency, and thus evidence of the AMH, of the futures contracts for the different products in the sample. Even if the results from the other tests reveal nonlinear serial dependence, it is better to know the degree of inefficiency and in what periods this happens, for forecasting and decision-making purposes, amongst others. Given the results of the number of windows in which nonlinear dependence, and thus AMH, is detected, which goes from 2.63% of the windows for Soybean Oil and Coffee, to 9.87% of the windows in the case of Eggs, the most volatile contract of the sample, some light has been shed towards the AMH on the futures contracts under study.

**Table 4**
**Hinich bicorrelation test results in moving non-overlapped time windows**

| | Corn | Wheat | Cocoa | Soybean oil | Sugar | Coffee | Feeder cattle | Eggs |
|---|---|---|---|---|---|---|---|---|
| AR(p) model | 2 | 2 | 4 | 2 | 1 | 2 | 2 | 1 |
| No. of Windows | 152 | 152 | 152 | 152 | 152 | 152 | 152 | 152 |
| No. of significant windows (%) | 6(3.95) | 6(3.95) | 6(3.95) | 4(2.63) | 6(3.95) | 4(2.63) | 6(3.95) | 15(9.87) |
| Dates of significant H windows | 07/08/94-08/16/94 | 03/27/96-05/03/96 | 01/09/96-02/15/96 | 12/26/96-02/03/97 | 11/30/95-01/08/96 | 06/02/00-07/11/00 | 12/13/94-01/19/95 | 06/27/95-08/03/95 |
| | 03/27/96-05/03/96 | 01/11/02-02/19/02 | 07/26/02-09/03/02 | 09/28/00-11/06/00 | 02/04/97-03/13/97 | 10/14/02-11/20/02 | 05/18/95-06/26/95 | 10/23/95-11/29/95 |
| | 06/29/01-08/07/01 | 12/31/02-02/06/03 | 07/15/03-08/21/03 | 06/05/03-07/14/03 | 12/02/98-01/08/99 | 11/21/02-12/30/02 | 11/07/00-12/14/00 | 07/23/96-08/29/96 |
| | 08/22/03-09/30/03 | 11/01/06-12/19/06 | 10/27/04-12/03/04 | 04/03/09-06/08/09 | 10/14/02-11/20/02 | 12/06/04-01/12/05 | 04/01/02-05/08/02 | 02/04/97-03/13/97 |
| | 10/03/06-11/09/06 | 08/13/07-09/19/07 | 06/20/05-07/27/05 | | 05/11/05 -06/17/05 | | 11/10/03-12/17/03 | 01/22/98-03/02/98 |
| | 08/26/09-10/02/09 | 07/06/10-08/12/10 | 03/23/09-04/29/09 | | 06/09/09-07/16/09 | | 12/06/04-01/12/05 | 07/26/99-09/01/99 |
| | | | | | | | | 10/12/99-11/18/99 |
| | | | | | | | | 11/07/00-12/14/00 |
| | | | | | | | | 03/05/01-04/11/01 |
| | | | | | | | | 05/22/01-06/28/01 |
| | | | | | | | | 06/29/01-08/07/01 |
| | | | | | | | | 05/09/02-06/17/02 |
| | | | | | | | | 10/14/02-11/20/02 |
| | | | | | | | | 03/19/03-04/25/03 |
| | | | | | | | | 09/08/08-10/15/08 |

**Conclusions**

In this paper we have investigated the adaptive market hypothesis (AMH) of eight agricultural commodity futures contracts. We applied a battery of nonlinear tests to the returns series and concluded that there was nonlinear serial dependence in all. Furthermore, we applied the Hinich portmanteau bicorrelation test (H-test) to analyze if the nonlinear serial dependence could be localized and if such phenomenon was persistent or triggered by a few periods of time. We identified the non-overlapped time windows in which such behavior occurred, ranging from 2.63% of the total number of windows, for the case of soybean oil to 9.87% for eggs. Thus, evidence of the AMH on the returns of these futures contracts was shed.

It would be interesting to study the determinants in the total number of significant windows for these products, as well as the economic, political or social triggers of the nonlinear serial dependence observed, in order to investigate the possibility of forecasting when the phenomena would occur.

**Acknowledgements**

Omar Rojas acknowledges partial support from the "Research Fund UP-2013" of Universidad Panamericana Guadalajara.